\documentclass[sigconf,natbib]{acmart}
\AtBeginDocument{%
  \providecommand\BibTeX{{%
    \normalfont B\kern-0.5em{\scshape i\kern-0.25em b}\kern-0.8em\TeX}}}

\copyrightyear{2023}
\acmYear{2023}
\setcopyright{acmlicensed}\acmConference[SIGIR '23]{Proceedings of the 46th International ACM SIGIR Conference on Research and Development in Information Retrieval}{July 23--27, 2023}{Taipei, Taiwan}
\acmBooktitle{Proceedings of the 46th International ACM SIGIR Conference on Research and Development in Information Retrieval (SIGIR '23), July 23--27, 2023, Taipei, Taiwan}
\acmPrice{15.00}
\acmDOI{10.1145/3539618.3592040}
\acmISBN{978-1-4503-9408-6/23/07}

\newcommand{\ie}{\emph{i.e., }}
\newcommand{\eg}{\emph{e.g., }}

\newcommand{\wrt}{\emph{w.r.t. }}

\usepackage{enumitem}
\usepackage{verbatim}
\usepackage[ruled]{algorithm2e}
\usepackage{multirow}
\usepackage{subfigure} 
\usepackage{ragged2e}
\usepackage{caption}
\usepackage{graphicx}
\usepackage[normalem]{ulem}
\useunder{\uline}{\ul}{}
\usepackage{amsmath}
\usepackage{setspace}
\usepackage{bm}
\usepackage{multirow}
\usepackage{longtable}
\usepackage{adjustbox}

\begin{document}

\title{Prediction then Correction: An Abductive Prediction Correction Method for Sequential Recommendation}

\author{Yulong Huang}
\authornote{Both authors contributed equally to this research.}
\email{huangyulong@mail.ustc.edu.cn}
% \email{zy2015@mail.ustc.edu.cn}
\affiliation{%
  \institution{University of Science and Technology of China}
  \city{Hefei}
  \state{Anhui}
  \country{China}
}

\author{Yang Zhang}
\email{zy2015@mail.ustc.edu.cn}
\authornotemark[1]
% \email{zy2015@mail.ustc.edu.cn}
\affiliation{%
  \institution{University of Science and Technology of China}
  \city{Hefei}
  \state{Anhui}
  \country{China}
}

\author{Qifan Wang}
\email{wqfcr@fb.com}
\affiliation{%
  \institution{Meta AI}
  \city{Menlo Park}
  \state{California}
  \country{USA}
}

\author{Chenxu Wang}
\email{wcx123@mail.ustc.edu.cn}
\affiliation{%
  \institution{University of Science and Technology of China}
  \city{Hefei}
  \state{Anhui}
  \country{China}
}

\author{Fuli Feng}
\authornote{Corresponding author.}
\email{fulifeng93@gmail.com}
\affiliation{%
  \institution{University of Science and Technology of China}
  \city{Hefei}
  \state{Anhui}
  \country{China}
}



\begin{abstract}

Sequential recommender models typically generate predictions in a single step during testing, without considering additional prediction correction to enhance performance as humans would. To improve the accuracy of these models, some researchers have attempted to simulate human analogical reasoning to correct predictions for testing data by drawing analogies with the prediction errors of similar training data. However, there are inherent gaps between testing and training data, which can make this approach unreliable.
To address this issue, we propose an \textit{Abductive Prediction Correction} (APC) framework for sequential recommendation. Our approach simulates abductive reasoning to correct predictions. Specifically, we design an abductive reasoning task that infers the most probable historical interactions from the future interactions predicted by a recommender, and minimizes the discrepancy between the inferred and true historical interactions to adjust the predictions.
We perform the abductive inference and adjustment using a reversed sequential model in the forward and backward propagation manner of neural networks. Our APC framework is applicable to various differentiable sequential recommender models. We implement it on three backbone models and demonstrate its effectiveness. We release the code at \url{https://github.com/zyang1580/APC}.

\end{abstract}

% <concept_id>10002951.10003317.10003347.10003350</concept_id>
% <concept_desc>Information systems~Recommender systems</concept_desc>

% \ccsdesc[500]{Information systems~Recommender systems}

%\keywords{Sequential Recommendation, Prediction Correction}


\vspace{-25pt}

\vspace{-25pt}
\maketitle
\vspace{-10pt}
\section{Introduction}
Sequential recommendation~\cite{past-future} involves predicting the next item based on a user's historical interaction sequence.
% In recent years, numerous sequential recommendation models have been developed and achieved substantial success in both academic and industrial domains~\cite{bert4rec,denSasrec,adaranker,sasrec}. However, these models employ a one-shot prediction approach during testing. 
% In recent years, numerous sequential recommendation models have been developed and achieved substantial success~\cite{bert4rec,denSasrec,adaranker,sasrec}. However, these models make prediction in a one-shot manner, \ie making recommendation decisions with a simple forward propagation of trained models.
In recent years, many sequential recommender models have been developed, and they have achieved significant success~\cite{bert4rec,denSasrec,adaranker,sasrec,xin2022rethinking}. However, these models typically make predictions in a one-shot manner, where recommendation decisions are made with a simple forward propagation of the trained models.
% That is, terminate the prediction process once recommendations are generated without any checking and correcting, although almost all models cannot guarantee to be completely free of making errors~\cite{UTEC,Rec-PC}.
 % irrespective of the recommendation quality.  
 % In contrast, humans would  further check and correct their results to enhance performance during joining testing, e.g., an examination. 
% Apparently, this decision procedure is different from us humans who typically further check and revise predictions especially on hard samples.
This decision procedure appears to differ from that of humans, who often engage in further checking and revision of predictions, particularly when dealing with difficult samples.
 % Nevertheless, recommender models are unlikely to be completely free of prediction errors~\cite{UTEC,Rec-PC}. 
  % Nevertheless, almost all models cannot guarantee to be completely free of prediction errors~\cite{UTEC,Rec-PC}.
To further enhance the quality of recommendations, it is crucial to incorporate the ability of \textit{prediction correction} into sequential recommendation.

\begin{figure}[t]
  \centering
\includegraphics[width=0.85\linewidth]{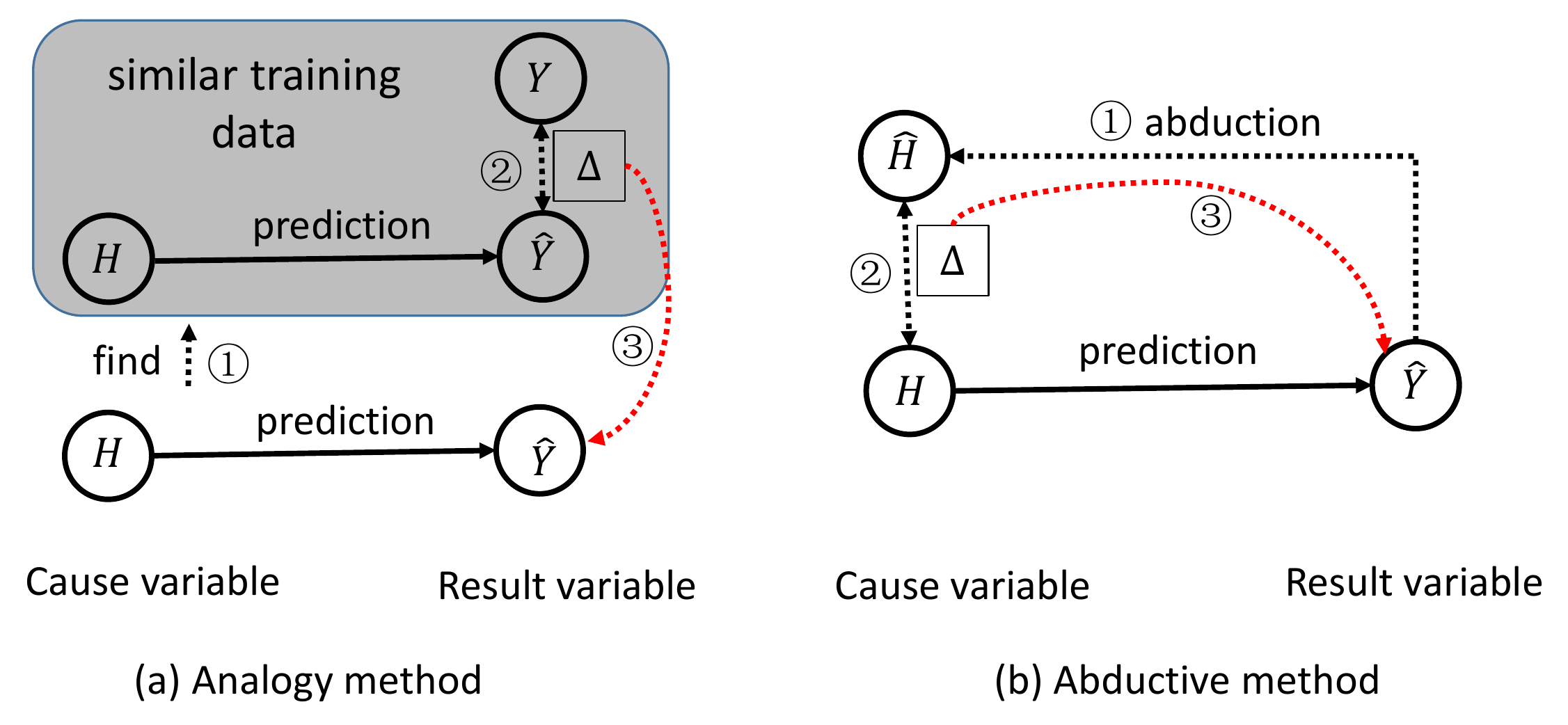}
%   \vspace{-10pt}
  % \caption{Illustration of prediction-correction methods for testing. $H$ ($\hat{H}$) denotes the (inferred) historical interactions/causes, and $Y$ ($\hat{Y}$) denotes the (predicted) future interactions/results. $\Delta$ denotes the difference. The real and dotted lines represent the prediction and prediction-correction processes, respectively. }
\vspace{-15pt}  \caption{Illustration of prediction correction methods. $H$ and $Y$ denote historical interactions (\ie causes) and future interactions (\ie results), respectively. $\hat{Y}$ is the model prediction, and $\hat{H}$ is the inferred historical interactions given the prediction $\hat{Y}$. $\Delta$ is the difference to guide the correction. The real and dotted lines represent the prediction and prediction correction processes, respectively.\vspace{-1mm}}
  \label{fig:method-cmp}
  \vspace{-5mm}
\end{figure}

Prior research has attempted to improve recommender model predictions by simulating one of the fundamental aspects of human thought: analogical reasoning~\cite{goswami2013analogical}. As illustrated in Figure~\ref{fig:method-cmp} (a), this approach involves retrieving similar training data for a testing candidate and adjusting the testing predictions by adding the prediction errors of such training data~\cite{UTEC,Rec-PC}.
However, we contend that this approach is unreliable to correct predictions for sequential recommender models due to inherent differences between the training and testing data~\cite{adaranker,Overfitting,zhang2020retrain,PDA}, including: 1) the model has already fitted the training data, but not the testing data; and 2) the training data pertains to the past, while the testing data pertains to the future. Such differences can lead to significant drifts in prediction errors between similar training and testing data, making it difficult to reliably correct testing predictions by analogy.

In addition to analogical reasoning, humans can also utilize abductive reasoning to evaluate their results, a key component of counterfactual thinking~\cite{pearl2009causality}. Abductive reasoning involves inferring the most plausible explanation, or causes, for the obtained results~\cite{abduction,AbductiveCS}.
As depicted in Figure~\ref{fig:method-cmp} (b), during testing, humans can use abductive reasoning to infer the causes for their results and then adjust those results by making the inferred and observed causes more consistent. Unlike the analogy-based approach, this method corrects the results by referencing the observed part of the testing instance rather than other instances. Therefore, simulating human abductive reasoning could be a promising approach to correct model predictions in sequential recommendation without relying on the prediction errors of training data.

To simulate human abductive reasoning for prediction correction in sequential recommendation, a key step is to design an appropriate abductive reasoning task. Since causes typically precede results, a user's past interactions and future interactions can be regarded as the cause and result in abductive reasoning, respectively. Based on this, we can formulate the abductive reasoning task as inferring the user's historical interactions from the predicted future interactions. The predictions can then be corrected by minimizing the discrepancy between the inferred historical interactions and the true historical interactions. However, a critical challenge is how to transfer the obtained discrepancy information to the predictions for correction effectively. Furthermore, there is a trade-off between the degree of correcting predictions and the risk of introducing errors when applying abductive reasoning. Thus, it is essential to strike a balance between the correction of the predictions and the reliability of the inference results in the abductive reasoning process.

To address these issues, we propose a novel \textit{Abductive Prediction Correction} (APC) framework, which includes a reversed sequential recommender model specifically designed for abductive reasoning tasks. Our framework corrects the predictions of a given recommender model through two key steps: abduction and adjustment. First, the APC framework uses the abductive model to infer the historical interactions that are most likely to have led to the predicted future interactions. Second, it adjusts the predictions of the original model by minimizing the difference between the inferred and true history using gradient descent. The difference information obtained from the abductive model is effectively utilized for correcting the predictions. To prevent over-correction, we only consider the top-$N^{\prime}$ ranked candidate items and reject the correction if it does not provide additional information gain for the abductive inference.

The main contributions of this work are summarized as follows:
\vspace{-5pt}
\begin{itemize}[leftmargin=*]
    \item We propose to simulate human abductive reasoning to correct recommender predictions for further enhancing performance.
    % \item We propose the APC framework for sequential recommendation, which minimizes the difference between abductively inferred and true historical interactions to correct model predictions.
    \item We propose the APC framework for sequential recommendation, which is applicable to differentiable sequential recommenders.
    % \item We conduct experiments with three sequential recommenders, verifying the superiority of our proposal over existing methods.
    \item We conduct experiments with three sequential recommender models, verifying the superiority of our proposal.
\end{itemize}

\vspace{-10pt}
\section{Related Work}
% We briefly summarize existing work on sequential recommendation and prediction correction.

% prediction correction
% recommender system
%% type1: 直接对于模型的结果就行纠正
%% paper1： Improving recommender systems via a Dual Training Error based Correction approach， 需要用到trianing阶段的ground truth label
%%paper 2: A User Training Error based Correction Approach combined with the Synthetic Coordinate Recommender System. UMAP
%% type2: 蒸馏模型中，
%% paper1： AAAI2022-Cross-Task Knowledge Distillation in Multi-Task Recommendation， 需要用到ground-truth label
%%
% beyond recommender system
%% Real-time Transportation Prediction Correction using Reconstruction Error in Deep Learning
%% Methods for correcting inference based on outcomes predicted by machine learning

% \noindent \textbf{Sequential recommendation.} Early sequential recommendation models focused on modeling transition information based on Markov Chain assumption \cite{rendle2010factorizing}. Later, neural networks like RNN \cite{GRU4Rec}, CNN \cite{caser,fajie}, Attention networks \cite{sasrec,past-future,bert4rec}, GNN \cite{gnn1,SGNN2019}, and MLP \cite{zhou2022filter} were used to better utilize sequential information. Contrastive learning has also been applied to sequential recommendation \cite{lin2022dual,xie2022contrastive}. However, none of these works have studied prediction correction. Our method differs from bidirectional models such as BERT4Rec \cite{bert4rec} and DualRec \cite{past-future} since we use future information to correct predictions during testing, whereas they only utilize future information during training.

\noindent \textbf{Sequential recommendation.} Early sequential recommendation methods focused on modeling transition information based on Markov Chain assumption \cite{rendle2010factorizing}. Later, many neural networks have been employed to 
better leverage sequential information,
% build sequential models for 
% better leveraging sequential information, 
such as RNN~\cite{GRU4Rec}, CNN~\cite{caser,fajie}, Attention networks~\cite{sasrec,past-future,bert4rec}, GNN~\cite{gnn1,SGNN2019} and MLP \cite{zhou2022filter}. Recently, contrastive learning has also been applied to sequential recommendation~\cite{lin2022dual,xie2022contrastive}. However, to our knowledge, none of these works have studied prediction correction. One may think that our method is similar to bidirectional models such as BERT4Rec~\cite{bert4rec} and DualRec~\cite{past-future}, as they both leverage future information for prediction. However, there are inherent differences: we leverage the information for correcting predictions during testing, while they are only during training.

% Early attempts at sequential recommendation mainly focused on modeling transition information between sequentially interacted items based on the Markov Chain assumption~\cite{rendle2010factorizing}. Later, many neural networks have been employed to build sequential models for better leveraging sequential information, such as RNN~\cite{GRU4Rec}, CNN~\cite{caser,fajie}, Attention networks~\cite{sasrec,past-future,bert4rec}, GNN~\cite{gnn1,SGNN2019} and MLP \cite{zhou2022filter}. Recently, contrastive learning has also been applied to sequential recommendation~\cite{lin2022dual,xie2022contrastive}. However, to our knowledge, no existing work has studied prediction correction. One may think that our method is similar to bidirectional models such as BERT4Rec~\cite{bert4rec} and DualRec~\cite{past-future}, as they both leverage future information for prediction. However, there are inherent differences: we leverage the information for correcting predictions during testing, while they are only during training.

\noindent \textbf{Prediction correction.} 
Prediction correction for machine learning models has gained increasing attention in various areas, \eg transportation prediction~\cite{transportation-PC}, medicine~\cite{wang2020methods}, 
% outcome prediction in medicine~\cite{wang2020methods}, 
trajectory prediction~\cite{trajectory-prediction}, and recommendation~\cite{UTEC, Rec-PC,KD-prediction-correct}. ~In recommendation, previous research can be divided into two lines. First, \cite{UTEC, Rec-PC} use analogy reasoning during testing to correct recommender predictions and enhance recommendation performance. Second, \cite{KD-prediction-correct} focuses on knowledge distillation with ground-truth labels. 
% Both lines of methods 
All these methods rely on the prediction errors of training data,
%focuses on knowledge distillation scenarios and corrects the teacher model's predictions by comparing them directly to ground-truth labels to improve the learning of student models during training. Both lines of methods rely on the prediction errors of training data.
while we avoid using such prediction errors by employing abductive reasoning.

% \subsection{Directly correct the model results} Prediction correction has 

% In Improving recommender systems via a Dual Training Error based correction approach, the author proposed a method to measure the error of recommendation from the perspective of items and users, and designed a dual system to combine the error of both, but this method needs to use the ground truth in the training stage, which will limit the model. In addition, A User Training Error based Correction Approach combined with the Synthetic Coordinate Recommender System is based on the user's training error correction, which is corrected according to the distance between the user and the object in the composite Euclidean coordinate system.

% \subsection{Distillation model}

% Cross-Task Knowledge Dissemination in Multi-Task Recommendation proposes a new correction mechanism to promote synchronous training of student model and teacher model. However, the disadvantage of this method is that it still needs to use the ground truth label of the training set.

% Similar methods are also widely used in other fields, such as Real-time Transportation Prediction Correction using Reconstruction Error in Deep Learning, which designs the reconstruction error of the model in the transportation system, and corrects the prediction. Methods for correcting inference based on outputs predicted by machine learning put forward the method of inferring after prediction in medical-related tasks, and corrected the deviation of the verification set using the rules observed in the test set.

\vspace{-5pt}
\section{Method}
Before presenting the proposed framework, we first introduce the problem, the basic concepts, and the notations in this paper.

\vspace{-10pt}
\subsection{Problem Formulation}
% Let $u \in \mathcal{U}$ and $v \in \mathcal{V}$ denote a user and an item, respectively. Each user corresponds to a chronological interaction sequence $S_{u}=[v_1,\dots,v_{t},\dots,v_T]$, where $v_{t}\in \mathcal{V}$ is the $t$-th  item interacted by $u$, $T$ represents the length of the sequence\footnote{If the amount of interacted items is less than $T$, we add some padding values in the left of the sequences.}. Let $\mathcal{D}=\{S_{u}|u \in \mathcal{U}\}$ denote the interaction sequences of all users. We build a sequential recommender model $f_{R}$ by fitting $\mathcal{D}$. During testing, given a sequence $S_{u}=[v_1,\dots,v_{t},\dots,v_T]$ for a user $u$, $f_{R}$ could generate a recommendation scores for each candidate item to represent how likely the item would be the $T+1$ interacted item. 
% In normal cases, the recommender  directly generates the final recommendation list according to the prediction scores. 
% Differently, we aim to
% further correct the predictions generated by $f_{A}$ to enhance recommendation performance.
Let $u \in \mathcal{U}$ and $v \in \mathcal{V}$ denote a user and an item, respectively, where $\mathcal{U}$ ($\mathcal{V}$) denotes the set of all users (items). Each user corresponds to a chronological interaction sequence $S_{u}=[v_1,\dots,v_{t},\dots,v_T]$, where $v_{t}\in \mathcal{V}$ is the $t$-th item interacted by $u$, and $T$ represents the length of the sequence\footnote{ We apply padding if the sequence contains fewer than $T$ interactions.
}. Let $\mathcal{D}=\{S_{u}|u \in \mathcal{U}\}$ denote the interaction sequences of all users. We build a sequential recommender model $f_{R}$ by fitting $\mathcal{D}$. During testing, given a sequence $S_{u}=[v_1,\dots,v_{t},\dots,v_T]$ for a user $u$, $f_{R}$ could generate a recommendation score for each candidate item, indicating how likely the item would be the $(T\text{+}1)$-th interacted item.
Normally, the recommender generates the final recommendation list based on the prediction scores.
Differently, we aim to further improve the recommendation performance by correcting the predictions generated by $f_{R}$.

% take a further step to correct model predictions to enhance recommendation performance.
% improve recommendation quality by  correcting model predictions. 
% taking a further prediction correction operation after a recommender model generates predictions. 
% Next, we present the proposed .
% \begin{figure}[t]
%   \centering
% \includegraphics[width=0.7\linewidth]{figs/framework_v3.pdf}
% \vspace{-15pt}
%   \caption{Overview of our abductive prediction correction framework. The real lines represent the initial prediction process. The dotted lines represent the correction process with two key steps: abduction inference and adjustment.}
%   \label{fig:framework}
%   \vspace{-8mm}
% \end{figure}
\begin{figure}[t]
  \centering
\includegraphics[height=3.5cm]{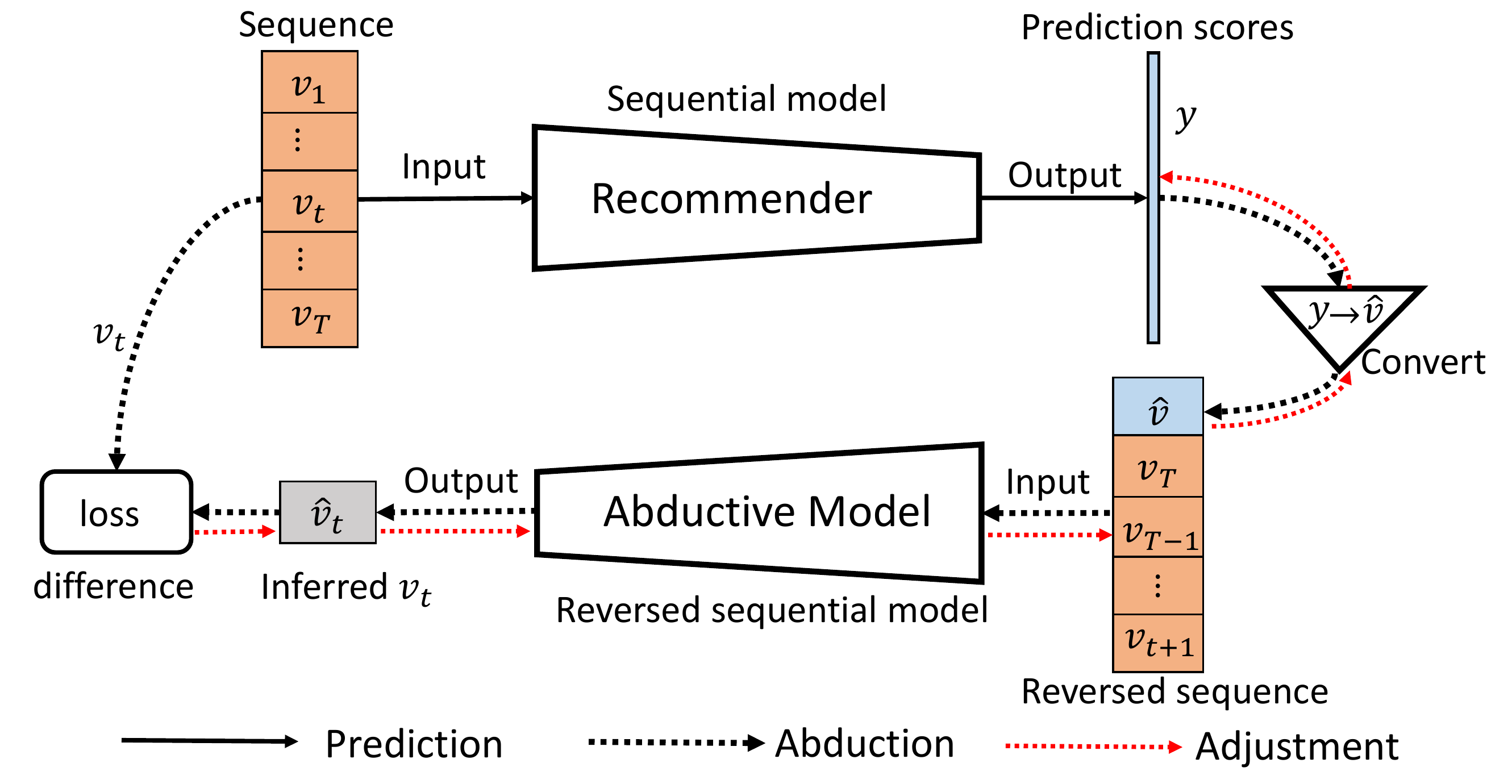}
\vspace{-15pt}
  \caption{Overview of our abductive prediction correction framework. The real lines represent the initial prediction process. The dotted lines represent the correction process with two key steps: abduction inference and adjustment.}
  \label{fig:framework}
  \vspace{-7mm}
\end{figure}

\vspace{-10pt}
\subsection{Abductive Prediction Correction Framework}
We describe prediction correction using abductive reasoning, including the framework and its components.
%We next present how to achieve prediction correction using abductive reasoning. Specifically, we first present the overall framework and then elaborate on the different components of the framework.
\vspace{-5pt}
\subsubsection{Overall Framework}
% We design a generic abduction-based prediction-correction (APC) framework, as illustrated in Figure~\ref{fig:framework}.  
% To simulate human abduction-based rethinking, our APC abductively infers the historical interaction based on the predicted next item and then corrects the prediction by minimizing the difference between the inferred and observed history.  

Inspired by human abductive reasoning, we develop a generic Abductive Prediction Correction (APC) framework for sequential recommendation (Figure~\ref{fig:framework}).
 The framework simulates abductive reasoning to construct a task of inferring the historical interactions based on the predictions of the recommender model $f_{R}$, and then corrects the predictions according to the inferred results.
To achieve this, apart from the recommender model $f_{R}$, we additionally build an abductive model $f_{A}$, a reversed sequential model, which could infer the past interactions given the future interactions. Then, we utilize the abductive model to correct the predictions generated by $f_{R}$ with two key steps: 
\begin{itemize}[leftmargin=*]
    \item[1)] abduction (black dotted lines in Figure~\ref{fig:framework}),  which infers the most possible historical interactions given the predicted scores of $f_{R}$,
    \item[2)] adjustment (red dotted lines in Figure~\ref{fig:framework}),
    which updates the prediction scores by minimizing the difference between inferred and observed historical interactions in a gradient descent manner.
    % updating the prediction scores by minimizing the difference between inferred and observed historical interactions in the gradient descent manner.
\end{itemize}

% 1) abduction, which infers the most possible historical interactions given the predicted scores of recommender models with the abduction model, and 2) update, which updates the prediction scores by minimizing the difference between inferred and observed historical interactions in the gradient update manner.
% \begin{itemize}[leftmargin=*]
%     \item Abduction, which infers the most possible historical interactions given the predicted scores of recommender models with the abduction model.  
%     \item Update, which updates the prediction scores by minimizing the difference between inferred and observed historical interactions in the gradient update manner.
% \end{itemize}
\noindent The two steps iterate until convergence or a maximum number of iterations is reached. Next, we present details for each key part.
% could be automatically corrected by performing the forward and backward propagation of the abductive model

\vspace{-5pt}
\subsubsection{Recommender and Abductive Models}
In our framework, the recommender model $f_{R}$ could be any well-trained embedding-based sequential recommender model, \eg SASRec~\cite{sasrec}.
That means, $f_{R}$ deals with the input interaction sequence by converting the interacted items to corresponding embeddings.
% To better learn the recommender model, we follow the corresponding training method taken by the original paper of proposing the model. 
Formally,  given a testing sequence $S_{u}=[v_1,\dots,v_{t},\dots,v_T]$, $f_{R}$ generates a recommendation (prediction) score $y_{v}\in [0,1]$ for each candidate item $v$ as:
% \begin{equation}\label{eq:prediction}
%     y_{v} = p_{R}(v|[v_{1},\dots,v_{T}]) = f_{R}(v;[{e}_{v_{1}},\dots,e_{v_{T}}]),
% \end{equation}
\begin{equation}\small\label{eq:prediction}
    y_{v} = f_{R}(v;[{e}_{v_{1}},\dots,e_{v_{T}}]),
\end{equation}
where $e_{v_{1}}$ denotes the learned item embedding for item $v_1$ in $f_{R}$, similarly for others. 
% Finally, $f_{R}$ generates a recommendation list $R_{u}$ for $u$ that contains the candidate items with the top $N$ highest prediction scores for $u$.
The abductive model $f_{A}$ is a reversed sequential model which infers the past interactions according  to the future interactions. To simplify, we keep the model architecture of $f_{A}$ the same as that of $f_{R}$. We train $f_{A}$ by fitting the reversed interaction sequences $\mathcal{D}^{-1}=\{S_{u^{\prime}}^{-1} | u^{\prime} \in \mathcal{U}\}$, where $S_{u^{\prime}}^{-1}$ is the reversed $S_{u^{\prime}}$ (\eg $S_{u}^{-1}=[v_{T},\dots,v_{1}]$),  and keep the training process\footnote{ We use the training method taken by the paper proposing the recommender model.} similar to that of the recommender model $f_{R}$.

% $f_{R}$ 
% $y_{v}$ could represent how likely the item $v$ will be interacted by $u$.

% it is trained to fit historical sequences $\mathcal{D}=\{S^1,\dots, S^{u},\dots, S^M\}$, where $S^{u}=\{v_{1}^{(u)},\dots,v_{n_{u}}^{(u)}\}$ denotes the historical sequence for user $u$. The abduction model is also a sequential recommender model with the same model architecture as the recommender model but is used to infer history instead of the next items. Specifically, it is trained to fitting the inverse historical sequences $\bar{\mathcal{D}} =\{\Bar{S}^1,\dots, \Bar{S}^{u},\dots, \Bar{S}^M\}$, where $\Bar{S}^{u}$ denotes the inverse of $S^{u}$. The detailed training method could refer to the paper proposed the recommender model, \eg SASRec~\cite{sasrec}. 
% We denote the well-trained recommender and abduction models as $f_{R}$ and $f_{A}$, respectively.
% %but just train  with the inverse sequences.
% During prediction, given a user with historical sequential interaction $S^{u}$, $f_{R}$ will encode the sequence with embedding sequence $[\mathbf{v}_{1}^{(u)},\dots,\mathbf{v}_{n_{u}}^{(u)}]$, and then generates a prediction score $y_{i}$ for each candidate item $v_{i}$, formally:
% $${y_{i}} = f_{R}(v_{i};[\mathbf{v}_{1}^{(u)},\dots,\mathbf{v}_{n_{u}}^{(u)}]),$$
% similarly for the abduction $f_{A}$. Finally, the recommender model $f_{R}$ would generate a recommendation list $R_{u}$ that contains candidate items with the top $N$ highest prediction scores for $u$. 

\vspace{-5pt}
\subsubsection{Prediction Correction} 
After the recommender model $f_{R}$ generates predictions scores for user $u$,
% After getting the predictions of the recommender model, 
we take two key steps to correct the prediction scores with the well-trained abductive model $f_{A}$:  
% To correct the recommendation results, we indeed need to correct the prediction scores of candidate items, as the scores  determine the ranking of candidate items.
% Next, we present the details of the abduction step and update step for correcting the prediction scores.

% To avoid introducing massive computation costs in the correcting process,  we only consider partial candidate items that are more important for a user. Specifically, considering that for each user $u$, we only correct the prediction scores of $R_{u}(K^{\prime})$, \ie the candidate items with the top $K^{\prime}$ ($K^{\prime}>K$) highest prediction scores for $u$.  

% Considering that the number of candidate items is very large,
% Specifically, to avoid massive computing overhead, we only consider update prediction scores for $R_{u}(K^{\prime})$, \ie candidate items with the top $K^{\prime}$ ($K^{\prime}>K$) highest prediction scores for $u$.  

%\vspace{+3pt}
\noindent \textbf{Step 1. Abduction}. 
In this step, we use the abductive model $f_{A}$ to infer the historical interactions with the prediction scores generated by $f_{R}$. The abductive model $f_{A}$ is an embedding-based sequential model that cannot directly utilize the prediction scores. 
To address the issue, we convert the prediction scores into a dummy item embedding, making them manageable by $f_{A}$.
Specifically, we generate the dummy item embedding $\hat{e}$ via embedding fusion with the prediction score-related weights as follows:
% we can easily enable $f_{A}$ to deal with prediction scores by converting them into a dummy item embedding $\hat{e}$ through embedding fusion. 
%Here, $\hat{e}$ could represent a dummy item $\hat{v}$ that is  predicted by $f_{R}$ the $T\text{+}1$ time point. 
% Specially, we generate the dummy embedding according to the prediction scores as follows:
\begin{equation}\small
    \hat{e} = \sum_{v\in \mathcal{V}^{\prime}_{u}} w_{v} e_{v},
\end{equation}
where $\mathcal{V}^{\prime}_{u}$ denotes the candidate items considered during the correction process for user $u$, $e_{v}$ is the learned embedding for item $v$ in $f_{A}$, and $w_{v}$ is a  prediction score-related weight to control the contribution of $e_{v}$ to $\hat{e}$. Formally, 
\begin{equation}\small\label{eq:weights}
w_{v} = \frac{(y_{v})^{\eta}}{\sum_{v^{\prime}\in \mathcal{V}^{\prime}_{u}} (y_{v^{\prime}})^{\eta} } ,     
\end{equation}
where $\eta > 0$ is a hyper-parameter to smooth the prediction scores, and $y_{v}$ is the prediction score generated by $f_{R}$ with Equation~\eqref{eq:prediction}.

Here, $\hat{e}$ could represent a dummy item $\hat{v}$ that $f_{R}$ predicts as the $(T\text{+}1)$-th interacted item for $u$.  We first inject $\hat{v}$ into the first position of the reversed interaction sequence of $u$. 
Then,  
% assuming that the $(T\text{+}1)$-th interacted item by $u$ would be $\hat{v}$, 
we take $f_{A}$ to  infer how likely the historical item $v_{t}$ ($t<T$) was interacted by $u$ at the $t$-th position, with the sub-sequence $[\hat{v},v_T, v_{T-1}, \dots, v_{t+1}]$ as model input. Formally, 
% the abduction model $f_{A}$ will infer how likely the $k$-th historical item $v_{k}$ is interacted by $u$ in the $k$-th point if the $t$-th interacted item is $v_{t}$. 
% Formally,
% the prediction that the $v_{k}$ is interacted in the $k$-th point of $S_{u}$ as follows:
\begin{equation}\small
   p_{A}(v_{t}|[\hat{{v}}, {v}_{T},\dots, {v}_{t+1}]) = f_{A}(v_{t};[\hat{e}, e_{v_{T}},\dots, e_{v_{t+1}}]), 
\end{equation}
where $p_{A}(v_{t}|[\hat{v}, {v}_{T},\dots, {v}_{t+1}])$ denotes the inferred probability, and $e_{v_{T}}$ denotes item embedding in $f_{A}$ for $v_{T}$, similarly for others. 
Next, we take the following loss $\ell$ to quantify how the inferred results match true historical observations:
\begin{equation}\small\label{eq:difference}
    \begin{split}
        \ell = &-log(p_{A}(v_{t}|[\hat{{v}}, {v}_{T},\dots, {v}_{t+1}])). %\\&- \lambda log(1-p_{A}(v_{t}^{\prime}|[\hat{v}, \bm{v}_{T},\dots, \bm{v}_{t+1}])),
    \end{split}
\end{equation}
% where $v_{k}^{\prime}$ denotes a randomly sampled item not interacted by $u$ at the time point $t$.
A small $\ell$ means less difference between the abductively inferred and true historical interactions.

\noindent\textbf{Step 2. Adjustment}. 
% After obtaining $\ell$ that quantifies the difference between the inferred and true history, 
We next minimize $\ell$, the difference between the inferred and true history, to update the prediction scores in the gradient descent manner. That is, we  treat the prediction scores $\{y_{v}|v\in \mathcal{V}^{\prime}_{u}\}$ as learnable parameters, and take the gradient descent method to update $\{y_{v}|v\in \mathcal{V}^{\prime}_{u}\}$ with the optimization goal of minimizing $\ell$. Formally,
\begin{equation}\small
    y_{v} \leftarrow y_{v} - \alpha \frac{\partial\ell}{\partial y_{v}},
\end{equation}
where $\frac{\partial\ell}{\partial y_{v}}$ denotes the gradient of $y_{v}$ \wrt $\ell$, and $\alpha$ refers to the learning rate to control the update step size. Obviously, the abduction and adjustment steps correspond to the forward and backward propagation of $f_{A}$, respectively. 
 %with the prediction scores as the learnable parameters. 
 We iteratively run the two steps until convergence or a maximum number of iterations is reached.

\vspace{-5pt}
\subsubsection{Preventing Over-correction} 
As the abductive model could also make mistakes, we take two strategies to avoid  over-correction: 

\noindent 
\textbf{Controlling $\mathcal{V}_{u}^{\prime}$.} Instead of correcting the predictions for all candidates, we only focus on correcting the candidate items with top-$N^{\prime}$   (initial) prediction scores. That means, 
\begin{equation}\small\label{eq:considered-item}
    \mathcal{V}_{u}^{\prime}=\{v|y_{v} \, \text{ranks in the top $N^{\prime}$ among  all candidate items} \}.
\end{equation}
% Note that we have $N^{\prime} > N$. 
Here, $N^{\prime}$ is greater than the length of the final recommender list.
Besides avoiding over-correction, this strategy could also help reduce computation costs, which is crucial for recommendation since there are usually efficiency requirements during online serving.

\noindent\textbf{Information gain-based strategy.} Another strategy is to reject the correction if the corrected results cannot bring information gain for inferring historical interaction, compared to only using the sequence without injecting $\hat{v}$, \ie $[v_{T},\dots,v_{t+1}]$. Specifically, after the last iteration is finished, we compute $\ell$ defined in Equation~\eqref{eq:difference} again and compare it with another $\ell^{\prime}$ gotten with $[v_{T},\dots,v_{t+1}]$. Formally, $\ell^{\prime}$ is computed similarly to $\ell$, having: $\ell^{\prime} = -log(p_{A}(v_{t}|[{v}_{T},\dots, {v}_{t+1}]))$,  
% \begin{equation}\label{eq:difference-cmp}
%     \begin{split}
%         \ell^{\prime} = &-log(p_{A}(v_{t}|[\bm{v}_{T},\dots, \bm{v}_{t+1}])), 
%        % - \lambda log(1-p_{A}(v_{t}^{\prime}|[\bm{v}_{T},\dots, \bm{v}_{t+1}])),
%     \end{split}
% \end{equation}
where $p_{A}(v_{t}|[{v}_{T},\dots, {v}_{t+1}])=f_{A}(v_{t};$
$ [e_{v_{T}},\dots,e_{v_{t+1}}])$ is the   probability --- how likely $v_{t}$ was the $t$-th interacted item of $u$ --- abductively inferred by $f_{A}$ without using $\hat{v}$. 
Then, we reject the prediction correction if $\ell>\ell^{\prime}$, and recover the initial prediction scores. 

\noindent\textit{ Recommendation}. 
% After finishing correcting prediction scores, we re-rank candidate items in $\mathcal{V}_{u}^{\prime}$ according to the corrected prediction scores and select top-$N$ ranked items as recommendations. 
After finishing the prediction correction, we re-rank the candidate items in $\mathcal{V}_{u}^{\prime}$ based on the corrected prediction scores, and select the top-$N$ ranked items as final recommendations.

\begin{table}[]
\centering
%\caption{Performance comparison between backbone models, DTEC, and our APC. `R@10'/`N@10' denotes Recall@10/NDCG@10. `Ori' denotes the backbone model without prediction correction, and `+DTEC'/`+APC' denotes applying DTEC/APC to the backbone model.
% `RI' denotes the relative improvement in Recall@10 with respect to `Ori'.
%}
\caption{Performance of backbone models, DTEC, and our APC compared. ‘R@10’/‘N@10’ is Recall@10/NDCG@10. ‘Ori’ is the backbone model without prediction correction, and ‘+DTEC’/‘+APC’ is with DTEC/APC applied.
}
\vspace{-10pt}
\label{tab:overall}
% \resizebox{0.5\textwidth}{!}{%
\begin{adjustbox}{width=0.70\width,center}
\begin{tabular}{c|c|cc|cc}
\hline
\multirow{2}{*}{Backbone} &  Dataset & \multicolumn{2}{c|}{Beauty}  & \multicolumn{2}{c}{ML1M}     \\ \cline{2-6} 
                         & Method   & R@10 & N@10 & R@10 & N@10 \\ \hline
\multirow{3}{*}{SASRec}  & Ori      & 0.0294    & 0.0157  & 0.2481   & 0.1284  \\
                         & +DTEC    & 0.0295    & 0.0158  & 0.2479   & 0.1283  \\
                         & +APC     & \textbf{0.0325}    & \textbf{0.0176}  & \textbf{0.2505}   & \textbf{0.1297}  \\ \hline
\multirow{3}{*}{DualRec} & Ori      & 0.0590    & 0.0359  & 0.2173    & 0.1214  \\
                         & +DTEC    & 0.0592    & 0.0360  & 0.2171    & 0.1215  \\
                         & +APC     & \textbf{0.0594}    & \textbf{0.0360}  & \textbf{0.2240}    & \textbf{0.1237}  \\ \hline
\multirow{3}{*}{Caser}   & Ori      & 0.0195    & 0.0098  & 0.2775    & 0.1469  \\
                         & +DTEC    & 0.0193    & 0.0097  & 0.2770   & 0.1468  \\
                         & +APC     & \textbf{0.0198}    & \textbf{0.0102}  & \textbf{0.2787}    & \textbf{0.1474}  \\ \hline
\end{tabular}%
\end{adjustbox}
% }
\vspace{-5mm}
\end{table}

\vspace{-10pt}
\section{Experiments}
We conduct experiments to verify the effectiveness of our proposal. 
% in achieving prediction correction for sequential recommendation.
%We conducted experiments to assess the effectiveness of our proposed approach in improving sequential recommendation accuracy through prediction correction.

\vspace{-10pt}
\subsection{Experimental Setting}
\noindent\textbf{Datasets.} We conduct experiments on two  representative datasets: Amazon-Beauty (Beauty)~\cite{mcauley2015image}, which  includes user reviews of products in Amazon, and MovieLens-1M (ML1M)~\cite{harper2015movielens}, which is a movie rating dataset collected by GroupLens Research\footnote{\url{https://grouplens.org/datasets/movielens/}.}. We preprocess the dataset following the setting in the SASRec work~\cite{sasrec}. Specifically, we discard users and items with fewer than five interactions, and  for each user, we select the most recent interaction for testing, the second most recent interaction for validation, and the remaining interactions for training.

\noindent\textbf{Compared methods.} To show the effectiveness of the proposed APC framework, we apply it to three representative sequential recommender models: 1) SASRec~\cite{sasrec}, which is a left-to-right self-attention model,  2) DualRec~\cite{past-future}, which is a bi-directional self-attention model, and 3) Caser~\cite{caser}, which is a CNN-based model. On the one hand, we directly compare with these backbone models to study whether our method could further enhance the performance of these models.  On the other hand, we compare APC with a SOTA analogy-based prediction correction method named DTEC~\cite{Rec-PC}. % We also implement DTEC for the three backbone models.

\noindent\textbf{Evaluation metrics and hyper-parameters}. To evaluate the top-N (N=10) recommendation performance, we take two metrics: Recall@N and NDCG@N, and compute them with the all-ranking protocol --- all non-interacted items are the candidates. For all methods, we search the learning rate, $L_2$ regularization coefficient, dropout ratio in [1e-2, $\dots$, 1e-4], [1e-1, 1e-2, $\dots$, 1e-6], and [0, 0.3, $\dots$, 0.7], respectively. For the special hyper-parameters of backbone models, we search them in the ranges provided in their paper.  For our APC, we set the $N^{\prime}$ in Equation~\eqref{eq:considered-item} as $50$, and search $\eta$ of Equation~\eqref{eq:weights} in [0.5, 1, 2, 3]. When computing $\ell$ in Equation~\eqref{eq:difference}, we take the  items  interacted by other users in the same mini-batch as negative items.

\vspace{-10pt}
\subsection{Results and Discussion}
\noindent\textbf{Overall performance}. 
We summarize the results in Table~\ref{tab:overall}, where we have the
following observations:
% Table~\ref{tab:overall} summarizes the recommendation performance of compared methods. From the table, we have the
% following observations:
% Table~\ref{tab:overall} summarizes the recommendation performance of compared methods. From the table, we have the
% following observations:
\vspace{-3pt}
\begin{itemize}[leftmargin=*]
    \item When applying our APC to the three different types of backbone models (SASRec, DualRec, and Caser), our APC consistently improved recommendation performance in all metrics on both datasets. These results demonstrate that conducting prediction correction can further enhance recommendation performance. 
    % \item DTEC, unlike APC, may not enhance recommendation performance and may even have a negative effect. DTEC improves user item predictions by comparing them to similar training items the user has engaged with. However, temporal shifts in sequential recommendation cause disparities between training and testing data, reducing the accuracy of the correction. The study finds that simulating abductive reasoning produces better results for correcting predictions in sequential recommendation.

     % In contrast to our APC, 
     %  the prediction correction method DTEC cannot consistently improve the recommendation performance and may even harm it. 
     %    % the prediction correction method DTEC is not consistently effective in improving recommendation performance and can even have a detrimental impact.
     % % DTEC involves analogical reasoning and corrects the predictions of candidate items for a user by drawing on prediction errors of similar training items that the user has interacted with. 
     % DTEC corrects the predictions of candidate items for a user by drawing analogies with the prediction errors of similar training items that the user has interacted with. 
     % However, inherent differences exist between training and testing data, particularly in sequential recommendation where temporal drifts exist, making its correction unreliable. The results confirm the superiority of simulating abductive reasoning to correct predictions in sequential recommendation.
    \item In contrast, DTEC cannot consistently improve the recommendation performance and may even harm it. 
     DTEC corrects the predictions of candidate items for a user by drawing analogies with the prediction errors of similar training items the user has engaged with. 
     However, inherent differences exist between training and testing data, particularly in sequential recommendation where temporal drifts exist, making its correction unreliable. The results confirm the superiority of simulating abductive reasoning to correct predictions in sequential recommendation.
     
     \item SASRec outperforms DualRec on ML1M, but DualRec performs better on Beauty. However, our method enhances the performance of DualRec on both datasets, indicating its distinctiveness from other bidirectional methods. It effectively utilizes the future-to-past information of sequences through abductive reasoning.

\end{itemize}
% xxxxxxxxxxxxxxxxxxxxxxxxxxxx#######x\\xxxxxxxxxxx##########xxxxxxxxxxxxxxxxx
\begin{figure}
\centering
\subfigure[\vspace{-10pt}\textbf{Recall@10}]{\includegraphics[width=0.22\textwidth]{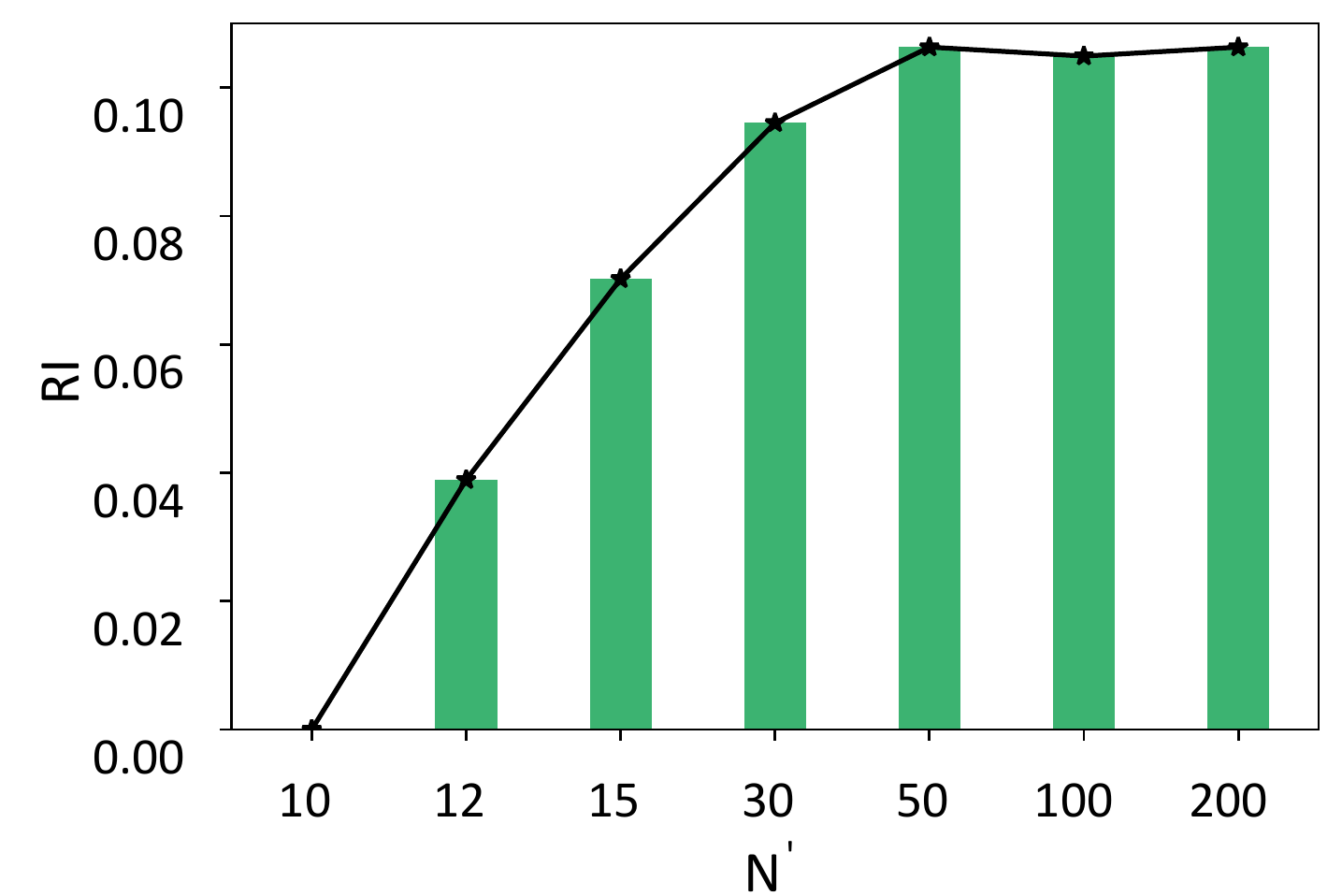}}
\subfigure[\vspace{-10pt}\textbf{NDCG@10}]{\label{fig:a}\includegraphics[width=0.22\textwidth]{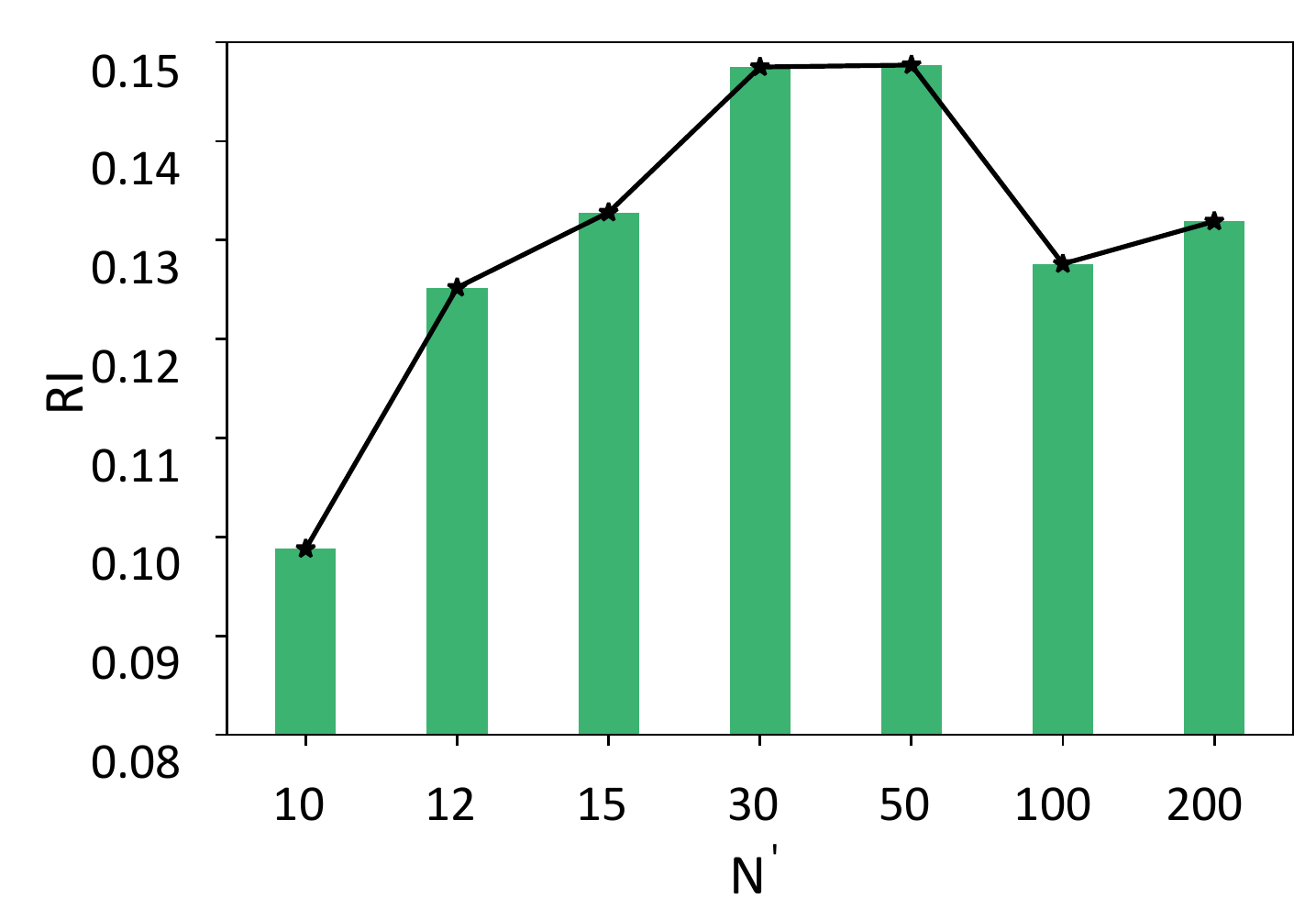}}
\vspace{-15pt}
\caption{
The effect of the size of $\mathcal{V}^{\prime}$ (\ie $N^{\prime}$) on the performance of APC. `RI' denotes the relative improvements of APC over SASRec on Beauty in corresponding metrics.
%Comparisons between different pruning thresholds on MF and LightGCN.
}
\vspace{-2mm}
\label{fig:considered-item}
\vspace{-18pt}
\end{figure}
\vspace{-3pt}
\noindent\textbf{In-depth analyses}. 
% Recall that we try to prevent over-correction by only considering items with the top-$N^{\prime}$ initial prediction scores, \ie $\mathcal{V}_{u}^{\prime}$, during prediction correction.  
% We next investigate how the size $N^{'}$ affects the relative improvements (RI) of our APC to backbone models on recommendation performance. Figure~\ref{fig:considered-item} summarizes the validation results on Beauty. Based on the figure, we can observe that as $N^{\prime}$ increases, the  RI in Recall@10 first increases and then slightly decreases, while the RI in NDCG@10 first increases and then decreases significantly. This suggests that a larger $N^{\prime}$ is beneficial to recall related items, but excessively large $N^{\prime}$ may reduce the ranking quality for related items. Therefore, it is crucial to choose an appropriate value of $N^{\prime}$ to avoid over-correction or under-correction. Besides, for the case $N^{\prime}=N=10$, the RI on NDCG is greater than $9\%$, this further verifies that APC could  improve ranking/recommendation quality for a fixed item list.   
To prevent over-correction, during prediction correction, we only consider the candidate items with top-$N^{\prime}$ initial prediction scores, \ie $\mathcal{V}_{u}^{\prime}$ in Equation~\eqref{eq:considered-item}. We next investigate how the size of $\mathcal{V}_{u}^{\prime}$ (\ie $N^{\prime}$) affects the relative improvements (RI) of our APC over backbone models in recommendation performance. Figure~\ref{fig:considered-item} summarizes the validation results on the Beauty dataset regarding the backbone model SASRec. We find that as $N^{\prime}$ increases, the RI in Recall@10 first increases and then slightly decreases, while the RI in NDCG@10 first increases and then decreases significantly. This suggests that a larger $N^{\prime}$ is beneficial for recalling related items, but an excessively large $N^{\prime}$ may reduce the ranking quality of related items. Therefore, it is crucial to choose an appropriate value of $N^{\prime}$ to avoid over-correction or under-correction. Additionally, when $N^{\prime}=N=10$, the RI in NDCG is greater than $9\%$, further verifying that APC can improve the ranking quality for a fixed item list. 
Besides, we find that $N^{\prime}$ has much greater impacts than the information gain-based strategy, so we omit the latter here.

\vspace{-10pt}
\section{Conclusion}

In this work, we introduce a universal APC framework for sequential recommendation inspired by human abductive reasoning. The framework formulates an abductive task of deducing the past interaction based on future interaction and minimizes the discrepancy between the inferred and actual history to correct predictions. We apply this framework to three representative sequential recommender models and validate its effectiveness.
In the future, we plan to extend the framework to other recommendation tasks such as collaborative filtering~\cite{ncf}. We also plan to leverage advanced techniques such as uncertainty~\cite{Uncertainty} to prevent over-correction.% to develop theoretically guaranteed strategies to prevent over-correction. %We will also explore other formulations for the abductive reasoning task in recommendation.
\vspace{-10pt}
\begin{acks}
This work is supported by the National Key Research and Development Program of China (2022YFB3104701), the National Natural Science Foundation of China (62272437,62121002), and the CCCD Key Lab of Ministry of Culture and Tourism.
\end{acks}

\bibliographystyle{ACM-Reference-Format}
\bibliography{7_reference}
\balance

% \appendix









\end{document}